\documentstyle[11pt,newpasp,twoside,epsf]{article}
\def\edcomment#1{\iffalse\marginpar{\raggedright\sl#1\/}\else\relax\fi}
\reversemarginpar

\begin{document}
\title{Disks surrounding Cataclysmic Binaries}
 \author{H.C.\ Spruit}
\affil{Max-Planck-Institut f\"ur Astrophysik, Postfach 1317, 85741 Garching, Germany}
\author{R.E.\ Taam}
\affil{Department of Physics \& Astronomy, Northwestern University, Evanston, IL 60208,
USA}

\begin{abstract}
Observations of Cataclysmic Variables show a number of phenomena that do not fit easily into
the standard magnetic braking scenario. These include the large spread in mass transfer rates,
the low surface temperatures of many of the companion stars, and evidence for material at low
velocities. We propose that these anomalies have a common cause: the presence of a circumbinary
(CB) disk. This may be a significant component of mass transferring binaries in general. Direct
detection of such CB disks may be possible but difficult, because of their low luminosity and 
spectral energy distribution peaking  in the mid-IR.
\end{abstract}

\section{Introduction}
The standard picture for the evolution of Cataclysmic Variables (CV), now nearly 2 decades old, 
has been reasonably successful as a framework for interpreting the phenomenology of CV.  In
this interpretation, angular momentum loss by gravitational waves and magnetic braking causes
cataclysmic binaries to evolve from longer to shorter orbital periods ($P$), with magnetic braking
dominating above the `period gap' ($P>3$ hr), and a lower braking rate below this gap ($P<2$
hr). This picture explains some overall statistical properties, for example the existence of the
period gap itself and the higher mass transfer rate above the gap compared to short period
systems. 

An additional element of the phenomenology, independent of the magnetic braking scenario but
not in conflict with it either, is the existence of outflows from the accretion disk. Evidence for
outflows comes from the profiles in UV-lines seen in outbursts of Dwarf Novae (e.g. Woods et
al. 1992),
possibly the single-peaked line profiles of the novalike (NL) and SW Sex binaries (see however
Hellier 2000 for an alternative explantation of these line profiles), and other indications for
outflow (e.g. Long et al. 1994).

\section{Anomalies}
A number of observations, however, are hard to fit into the standard picture:

\begin{itemize}
\item{1} A major problem is the very large spread in inferred mas transfer rates at a given
orbital period (Patterson 1984; Warner 1987). The braking scenario predicts a close
relation between mass transfer rate and orbital period.

\item{2} An extreme form of the spread in mass transfer rate is shown by the short-period super
soft sources (SSS). Though some SSS have long orbital periods and their large transfer rates can
be explained as the result of expansion of the secondary by nuclear evolution, several have short
orbital periods, in the same range as the novalike systems. The secondaries at such periods are
too small to have significantly evolved. An unknown process causes these binaries to transfer
mass at rates sufficient ($\sim 3\,10^{-7}{\rm M}_\odot{\rm yr}^{-1}$) to sustain the observed
steady burning of Hydrogen at the surface of the White Dwarf primary star.
  
\item{3} A second unexpected fact is the distribution of the novalike systems, which cluster at
periods (around 4-6 hr) just above the period gap, whereas the braking scenario predicts them
to predominate at longer periods, where the predicted mass transfer rate are highest. 

\item{4} The single peaked profiles of the emission lines in the SSS, the SW Sex stars, the AM
CVn stars and many NL systems show that in some cases the emission lines are not
predominantly formed in the accretion disk, but in optically thin material of large emission
measure somewhere else in the system. The phenomenon is particularly striking in some AM CVn
systems, (cf.\ Warner 1995), where a narrow  emission line is seen on top of a broad double
peaked emission line from the accretion disk. This phenomenon has been interpreted in terms
of circumbinary material already by Solheim and Sion (1994). A similar phenomenon may be the
low-velocity emission peaks in supersoft sources like QR And (Deufel et al. 1999).

\item{5} `Additional light' in systems with parameters such that mid-eclipse of the primary star
and accretion disk should be total. In some of these systems the light remaining at mid-eclipse 
can not be explained by the spectral type of the secondary star.  In the case of UU Aqr the 
spectrum of the additional light has been determined; it shows the characteristics of optically thin 
emission (Baptista et  al. 1998).

\item{6} Comparison of the spectral types of the secondary stars with theoretical models (Baraffe
and Kolb 2000) shows them to be significantly cooler, on average, than main sequence stars, as if
they were less massive and overexpanded as a result of a recent prolonged period of large mass
transfer.

\item{7} Superhumps above the period gap. The so-called superhumps, a feature in the lightcurve
that drifts in orbital phase, is thought to be due to a resonance in the accretion disk. This 
resonance can only occur at low mass ratios, such as occur in the short-period systems. They are 
also seen in several novalike systems, however (Patterson, 1999). This would be another 
indication that the mass of the secondary in such systems above the period gap is often 
smaller than expected from the standard braking scenario.

\item{8} Pile up at the minimum period.
The standard braking scenario predicts that the orbital evolution of CVs slows down as they
approach the minimum orbital period,
after which their orbital period increases again at a very slow rate. Most old CVs are
predicted to end up somewhere near this minimum period. The observed period distribution
does not show such a spike in the distribution (cf.\ Kolb \& Baraffe 1999). Though the
observations are probably biased against detection of these systems because of their low
quiescent luminosity and the long recurrence time of their outbursts, the discrepancy is large, and
can probably not be explained entirely by these selection effects (Kolb 2001; Marsh et al. 2001;
and references therein; G\"ansicke et al. 2001).

\item{9} Value of the minimum period. The theoretically predicted value of the minimum orbital
period of CVs (excepting thiose with helium star companions) is around P=70 min, the observed
value around 80 min. Stellar models are believed to be sufficiently accurate, but agreement with
observations would be obtained if the mean mass transfer rate of the short-period systems 
were at least a factor 3 higher than assumed in the standard braking scenario.

\item{10} The rate of magnetic braking. The standard formulas used for angular momentum loss
by  magnetic stellar winds is somewhat of an extrapolation, based on loss rates inferred from
stars rotating much more slowly than the secondaries of CVs. Recent work by Andronov,
Pinsonneault and Sills (2001)  indicates that this extrapolation overestimates the angular
momentum loss rate in CVs by as much as 2 orders of magnitude. If this is correct, the observed
mass transfer rates in CVs can not be due to magnetic braking.

\end{itemize}

\section{Interpretation}

Though some of the anomalies listed are possibly red herrings which may find unrelated
explanations, they all fit qualitatively into the suggestion made here, a CB disk, i.e. a disk
surrounding the binary. If these exist, they would be a major, thus far unrecognized 
component of Cataclysmic binary systems. 

\begin{figure}
\plotone{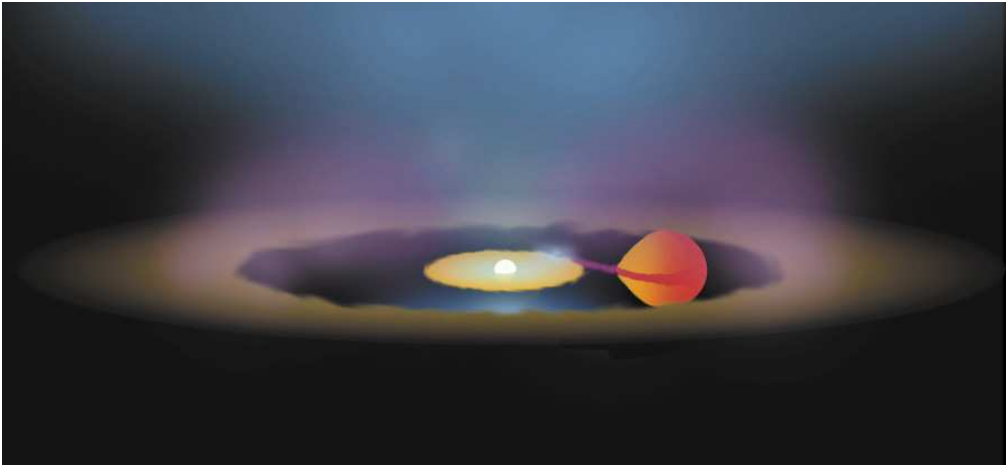}
%\plotone{cbdiskbw.ps}
\caption{Artist's impression of a circumbinary disk around a cataclysmic binary. The haze above
the disk indicates the outflow from the accretion disk. A slow component of the outflow
condenses onto the CB disk and feeds it.}
\end{figure}

A CB disk would be an additional source of angular momentum loss for the binary. The tides
raised in the CB disk by the orbiting binary transfers angular momentum to it, and prevent it
from spreading inward. Depending on the surface density of the CB disk, the angular momentum
loss from the binary would differ, and with it the mass transfer rate. The SSS would, in this
interpretation, be the relatives (Greiner et al. 2000) of NL and SW Sex systems, the difference
being the presence of a more massive CB disk.

The increased mass transfer rates caused by CB disks would be systematic, since a CB disk
would dissipate only very slowly by spreading. This might explain the apparent overexpansion of
the secondary stars. A higher mass transfer rate would also help to explain the discrepancy
between the observed and predicted values of the period minimum. 

The obvious question is how a CB disk would be formed. One possibility would be that it is a 
`fossil': a remnant from a late stage of the common envelope (CE) evolution phase in which the 
cataclysmic binary was formed (for a review see Taam \& Sandquist 2000). This could happen 
as a `fall back' process at the time the common envelope becomes optically thin. Given the 
difficulty of computing the CE evolution, it is hard to estimate the likelihood of such a formation 
channel for CB disks. 

A second possibility is that CB disks form as a by-product of  mass loss from the system. The
single peaked line profiles in SW Sex stars indicate outflow velocities up to 2000 km/s, which is
well above the escape speed from the binary. There is large optical depth in these lines at lower
velocities as well, however, so that a significant outflow component at velocities of the order of
the escape speed from the binary (a few 100 km/s) or lower may also exist. This would be
material that would `hang around' the binary rather than being ejected. It would naturally form a
CB disk by cooling and settling to the equatorial plane. It is hard to quantify from the
observations how much material might condense into a CB disk in this way. Instead, we have
used model calculations in which this uncertainty is parametrized.  These show that a relatively 
minor amount would suffice to build up substantial CB disk. 

\section{Model calculations}
For the model calculations (Spruit and Taam 2001, Taam and Spruit 2001, Dubus et al. 2001) we
have made use of the observation that evidence for outflows is strongest in systems with high
mass transfer rates: dwarf novae in outburst, the SW Sex stars, supersoft sources and the AM
CVn stars. This suggests that mass loss by outflows increases with mass transfer rate. We have
made the simple assumption that the two are proportional. In addition, we have assumed that of
the outflow, a fixed fraction settles into a CB disk. The mass input rate $\dot M_{\rm CB}$ into
the CB disk is thus a given fraction $\delta$ of the mass transfer rate $\dot M$. 

\subsection{Feedback}
The consequence of the assumed proportionality between mass input into the CB disk and the
mass transfer rate is a positive feedback: once a CB disk is present, the angular momentum it
extracts from the binary enhances the mass transfer rate from the secondary to the primary. This
increases the mass input rate into the CB disk, increasing its surface density. With this higher
density the increasing torque on the binary in turn increases the mass transfer rate. Though the
mass transfer might initially be caused by the standard magnetic braking process alone, the mass
building up in the CB disk would eventually make this feedback strong enough to become
unstable. From this point on, the mass transfer rate and the input rate into the CB disk would
increase in step exponentially. In this way the large mass transfer rates in for example the
short-period supersoft sources could develop. In our interpretation, these would be descendants
of novalike systems, as suggested before by Greiner et al. (2000). 

We call the phase of high mass transfer resulting from this instability `accelerated mass transfer'
induced by the CB disk. The calculations show that the orbital period typically `bounces' shortly
after accelerated mass transfer sets in (if the secondary is not significantly evolved) . The more
rapid mass loss from the secondary causes it to expand, forcing the orbital period to increase.
This contrasts with the evolution  under standard braking, which results in an essentially
monotonically decreasing orbital period.

Whether the effect of a CB disk on the binary becomes significant depends mostly on the mass 
input rate into it. Analytic results based on the above assumptions (Spruit and Taam 2001) 
suggested that a minimum value of  $\delta$ around $10^{-3}$ would be sufficient to cause 
accelerated mass transfer to occur. The precise value is unfortunately somewhat sensitive to the 
physics determining the viscous spreading of the disk. Remarkably, the uncertain value of the 
viscosity itself, as parametrized through a constant $\alpha$-viscosity, has only a weak influence. 
This is because the torque exerted by the disk is proportional to $\alpha\Sigma_{\rm i}$, where 
$\Sigma_{\rm i}$ is the surface density at the inner edge of the disk. A larger value of $\alpha$ 
is compensated largely by the more rapid spreading of the disk, which lowers $\Sigma_{\rm i}$. 
More important than $\alpha$ itself is the relative variation of the viscosity through the disk.
Calculations using more detailed physics for the CB disk (Taam and Spruit 2001) indicated that a
rather high value of $\delta$, around $10^{-2}$, would be needed for accelerated mass transfer, 
but these did not include the effect of
convection on the structure of more massive CB disks. This was included by Dubus et al.
(2001), together with a more sophisticated method for calculating the viscous evolution of the CB
disk. In these so far most realistic calculations values of $\delta$ around $10^{-4}$ were
found to be sufficient to cause a rapid acceleration to high transfer rates to happen within a
Hubble time.

%waiting time and acceleration time. analytic and Dubus etal. xxx
 
\section{Consequences of the presence of CB disks}

\subsection{Dissolution of the secondary}
In the standard magnetic plus gravitational radiation braking scenario, the mass transfer rate
declines rapidly with decreasing secondary mass, such that complete transfer of the secondary
mass does not happen within a Hubble time. With torques due to a CB disk the evolution of the
secondary is more interesting. In the `accelerated' transfer phase, the angular momentum loss is
determined by the surface density of the CB disk, rather than by the secondary star. Hence the
mass transfer rate stays finite as the secondary mass vanishes. As a consequence, the secondary
dissolves completely within a finite, relatively short time once a phase of accelerated mass
transfer has set in, and leaves a single white dwarf. 

This would have significant effects on the distribution of systems with orbital period. If a
substantial fraction of all systems go through a CB-induced phase of accelerated mass transfer,
they would disappear from the CV population, and the pile-up of systems towards the period
minimum around 80 min. would be replaced by a gradual decline of the number of systems. This
would be more in line with current observations (cf Kolb 2001).

\subsection{White dwarfs with planets?}
After dissolution of the companion, the white dwarf primary would be left as a `white widow',
with a CB disk still surrounding it. The mass in this disk could be substantial, perhaps as much as
$10^{-3}-10^{-2}\,{\rm M}_\odot$. As this disk cools, it may well form planets in the same 
way as a protoplanetary disk. It remains to be seen whether such planets would be detectable,
for example through Doppler shifts of the NLTE cores of the Balmer lines of the WD.

\subsection{The period gap}
The CB interpretation given here implies a large systematic spread in mass transfer rates. [As
opposed to cyclic mass transfer variations such as proposed by King et al. (1995). In this
mechanism the average mass transfer rate does not spread much]. Though this spread explains a
number of  facts such as the observed varying degree of overexpansion of the secondaries, it is
not easily compatible with the explanation of the period gap in terms of disrupted magnetic
braking. The period gap gets `washed out' if the long-term transfer rate differs too much
between systems. It remains to be seen whether a plausible alternative
explanation of the gap can be found within the CB scenario.

\section{Observability}
One may wonder how it is possible that something as serious as circumbinary disks around CVs
could have escaped detection so far. On closer inspection, however, it turns out that such disks
would in fact be rather difficult to observe. On the one hand, the luminosity of a CB disk is low.
Though it extracts a lot of angular momentum, this happens at a large distance in a shallow
gravitational potential, and is thus accompanied by little energy dissipation. This makes them
inconspicuous against the very bright accretion disk, which processes material deep in the
potential well of the primary. Secondly, the size of the CB disk is such that its spectral energy
distribution peaks in the poorly accessible mid IR (at 3-10$\mu$m, cf. the models of
Dubus et al., 2001). Finally, model calculations show that the CB disk probably lies in the shadow
of the outer part of the accretion disk. Enhanced brightness through illumination of the disk by
the bright central regions of the accretion disk is thus also unlikely.

\end{document}